\documentclass[reprint, amsmath, amssymb, aps]{revtex4-1}

\usepackage{amssymb}
\usepackage{amsmath}
\usepackage{float}
\usepackage{graphicx}
\usepackage{epstopdf}
\usepackage{dcolumn}% Align table columns on decimal point
\usepackage{bm}
\usepackage{setspace}
\usepackage{todonotes}
\usepackage{natbib}

\usepackage{hyperref}
\hypersetup{colorlinks=true, citecolor=blue, urlcolor=blue, linkcolor=blue}

\begin{document}
%\doublespacing
\title{Controlled transitions between phyllotactic states of repulsive particles confined on the surface of a cylinder}

\author{A. A. Tomlinson}
\author{N. K. Wilkin}
\affiliation{School of Physics and Astronomy, University of Birmingham, Edgbaston, B15 2TT, UK}

\date{\today}

\begin{abstract}
Phyllotactic states are regular lattice-like structures on cylinders and are a botanical classification scheme. In this communication, we report a sequence of transitions between phyllotactic states for particles with a repulsive particle-particle interaction on a cylindrical geometry at zero temperature. We can infer the transition points as a function of density via Monte Carlo simulations, as well as the mathematical descriptions of the ground states. The lattices we generate are described as phyllotactic states that fit onto the cylindrical surface as a set of helical chains. Our analysis shows how all state energies lie on the same parabola which we exploit to find the transitions.
\end{abstract}

%\pacs{45.70.-n}{Classical mechanics of granular systems}
%\pacs{87.18.Hf}{Pattern formation in cellular populations}
%\pacs{64.75.Yz}{Self-assembly}

\maketitle

%%%%%%%%%%%%%%%%%%%%%%
%% INTRODUCTION  %%  	  
%%%%%%%%%%%%%%%%%%%%%%
\section{Introduction}
In this communication, we propose an algorithm that enables us to analytically construct the infinite sequence of transitions between phyllotactic states of repulsive particles confined to the {\em surface} of a cylinder. We build upon research on the close-packing of spheres {\em inside} cylindrical tubes \cite{lohr2010,mughal2011,mughal2017,erickson1973} and disks in narrow channels \cite{godfrey2014,kofke1993}, minimum energy structures of nano-particles in carbon nanotubes \cite{khlobystov2004,liang2014,yamazaki2008}, and the arrangement of some areoles on cacti. All of these systems share the common mathematical description of phyllotaxis.

\begin{figure}
\centering
\includegraphics[width=8.6cm]{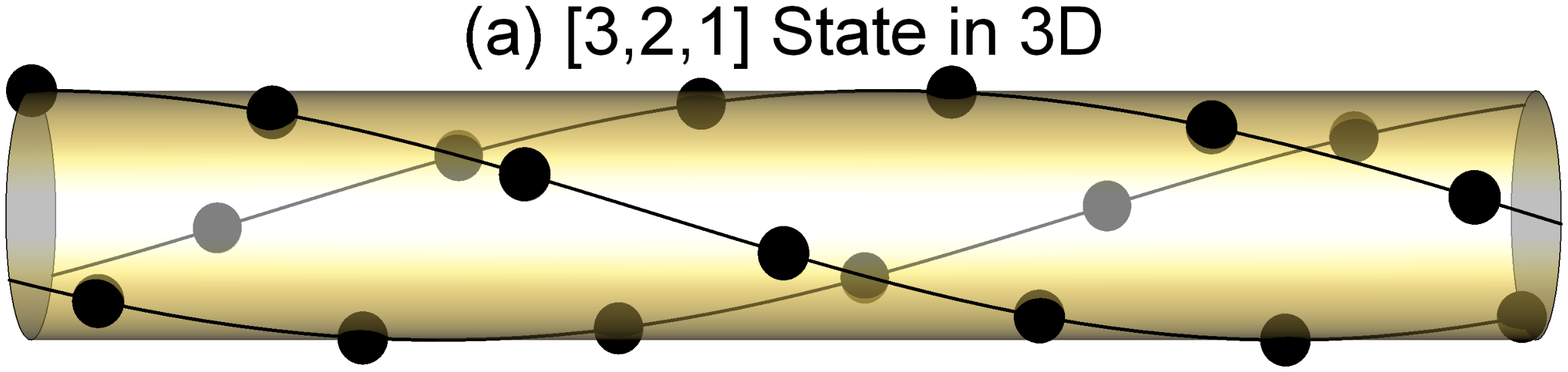}\\ 
\vspace{0.15cm}
\includegraphics[width=8.6cm]{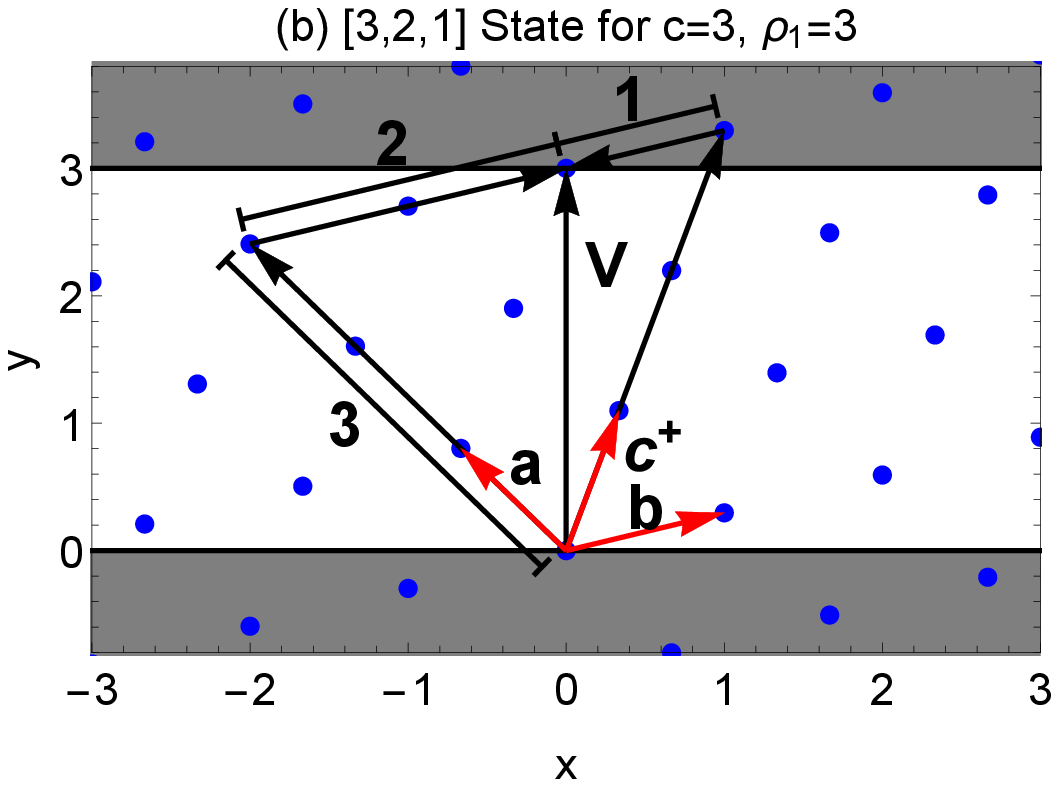}
\caption{(Color online) (a) Three-dimensional rendering of a \([3,2,1]\) ground state on a cylindrical segment when \(c=3\) and \(\rho_1=3\). (b) The same state is projected into two dimensions where we see the origin of the phyllotactic notation. The periodicity vector, \(\mathbf{V}=\mathbf{\hat{y}}c=m\mathbf{a}+n\mathbf{b}\), has a length equal to the circumference and is a combination of lattice vectors \cite{mughal2013}. The thick horizontal lines represent the periodic boundary condition. In this example, \(\mathbf{V}=3\mathbf{a}+2\mathbf{b}\).}
\label{fig:examples}
\end{figure}
Phyllotaxis is a biological classification used to described the pattern formation of leaves on plant stems \cite{airy1872}. The phyllotactic notation treats the cylindrical stem as an ``unravelled'' two-dimensional structure with a triangular lattice (formed by two lattice vectors) with node sites corresponding to petiole (stalk) locations \cite{mughal2013}. This can then be described as a set of N chains formed by following \(m\) and \(n\) multiples of the two lattice vectors to wrap once round the cylinder to the original node. The notation is expressed as \([m+n,m,n]\) or \([m,n,m-n]\), where \(m,n\in\mathbb{N}\). We distinguish these two representations because \(m\) and \(n\) generate one of the two phyllotactic structures depending on the system geometry and density of particles.

Optimal packing of spheres in cylinders and hard disks in periodic geometries are purely geometric problems. In three dimensions, shell-like structures of surface or core spheres form and are well described by phyllotaxis \cite{fu2016}. Similarly, hard disks form phyllotactic structures because of the periodicity vector defined in Fig. \ref{fig:examples}. The ground states of these systems can include helical grain boundaries (line-slips) \cite{nelson2016}. Line-slips can appear due to geometrical constraints which are not present in our soft-matter-type potential.

Related work \cite{piacente2004} on parabolic confinement demonstrates the number of rows of repulsive particles that form become dependent on the background potential and interparticle interactions. Transitions between the number of rows of parabolically confined particles follow similar trends to the helical row transitions that we report despite the differing boundary conditions.

Interactions may be a function of the three-dimensional distance between particles \cite{oguz2011}. Alternatively, enforcing periodic boundary conditions in one direction causes interactions to depend on the arc lengths on the surface of the cylinder. Our study is the latter choice, although our results show qualitative similarity for both interaction distances.

We begin by numerically determining the energetics of the ground state lattices close to the structural transition points. The analysis leads to the conjecture of a scale invariant, \(\alpha=c^2\rho_2\), where \(c\) is the circumference of the cylinder and \(\rho_2\) is two-dimensional particle density. The ground state is exclusively determined by \(\alpha\) - and this is verified by our numerical results.
%%%%%%%%%%%%%%%%%%%%%%
%%  	SYSTEM %%
%%%%%%%%%%%%%%%%%%%%%%
\section{System}
We model the cylinder as a two-dimensional planar system of repulsive particles with a periodic boundary condition imposed in the circumferential (\(y\)) direction. This model has been used previously for studying particle interactions on cylinders \cite{amir2013}. In the numerical simulations, the length of the cylinder is set to be much greater than the circumference, \(c\), to approximate an infinitely long cylinder by imposing an additional periodic boundary condition in the axial (\(x\)) direction, modelling the cylinder as a torus. We study the bulk behaviour of the numerical system to make phenomenological inferences in our analytical model which assumes an infinitely long cylinder. Structural ground states that form in the bulk when \(L\gtrsim10c\) are well-explained in the analysis. This is a zero temperature system.

To establish that the behaviour is not special to a particular interaction potential, we have worked with both a modified Bessel-function of the second kind, \(K_0(r/\lambda)\), and the Yukawa-potential, \(\exp(-\kappa r)/r\), where \(r\) denotes particle separation. These potentials are representative of two-dimensional soft matter system namely vortices in superconductors and Wigner Crystals. Qualitative similarities imply a generality to our results. Numerical results in this letter set \(\kappa=1\) and \(\lambda=1\) (arbitrary units).
%%%%%%%%%%%%%%%%%%%%%%
%%  Observations  %%
%%%%%%%%%%%%%%%%%%%%%%
\section{Initial numerical results}
We use a Metropolis Prescription of the Monte Carlo algorithm to anneal the system to zero temperature. Circumference and linear density, \(\rho_1\) (number of particles per horizontal length along the cylinder), are varied and an initial phase diagram is generated. The linear density is appropriate since the two-dimensional density can be scaled out using \(\rho_2=\rho_1/c\).

Figure \ref{fig:examples} is a typical observed state. Initial results indicate that the bulk of the system forms an isosceles (or equilateral in special cases) triangular lattice structure. The lattice is regular and unit cell sizes remain constant.

Because the system is numerically finite, commensurability effects can cause highly localised grain boundaries or lattice defects to form. This is due to helically defined states being unable to align correctly with themselves, so a discontinuity forms. However, the bulk of the system remains homogeneous which is where we derive our phenomenology.

Simulating between \(0.5\le c\le 5\) and \(0.5 \le \rho_1\le5\) with a resolution of \(0.1\times0.1\) shows that transition lines between states take the form \(c=\alpha_T/\rho_1\), where \(\alpha_T\) is a number to be determined for each transition. Figure \ref{fig:phasespace} shows the phase diagram generated by our subsequent analysis, which is qualitatively similar to these initial results.
%%%%%%%%%%%%%%%%%%%%%%
%% Lattice Model  %%
%%%%%%%%%%%%%%%%%%%%%%
\begin{figure}
\centering
\includegraphics[width=8.6cm]{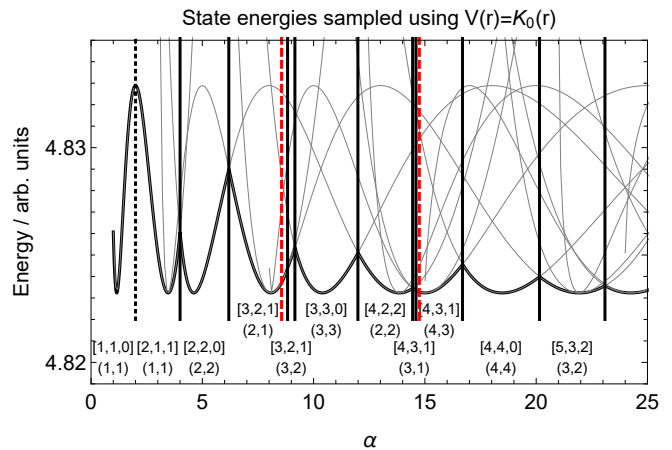}
\caption{(Color online) Energy curves for the Bessel function potential. Transitions are indicated by thick vertical lines for clarity. Each state is labelled in accordance to the phyllotactic notation (above) and the indices from the system model (below). When compared with the Yukawa potential, most transitions lines are nearly identical (the energy values are different, but appear to map to one another). The major differences with the Yukawa potential are shown in thick dashed red for the transitions: \([3,2,1]\) \((m=2,n=1)\rightarrow [3,2,1]\) \((m=3,n=2)\) and \([4,3,1]\) \((m=3,n=1) \rightarrow [4,3,1]\) \((m=4,n=3)\). The thin gray lines show the energies of candidate states not in the ground state. The dotted transition line at \(\alpha=2\) shows a `relabelling' transition which is discussed later.}
\label{fig:sampling}
\end{figure}
\section{Phenomenological model}
The three key observations from the numerical results we can build a phenomenological model with are:

\(\bullet\) constant unit cell size,

\(\bullet\) the periodicity of the lattice in the vertical direction,

\(\bullet\) and an isosceles lattice.\\
We combine these features to create a phenomenological model in terms of the lattice vectors.

TABLE I. For a regular isosceles lattice under periodic confinement in the \(y\) direction, the lattice vectors \(\mathbf{a}\) and \(\mathbf{b}\) are constrained as below,
\begin{center}
\begin{tabular}{|c|c|}
\hline
\textbf{Constraint} & \textbf{Physical Reason}\\
\hline
\(\mathbf{\hat{z}}\cdot(\mathbf{b}\times\mathbf{a})=c/\rho_1\) & Constant unit cell size\\
\hline
\(m\mathbf{a}+n\mathbf{b}=\mathbf{\hat{y}}c\) &  Periodicity vector\\
\hline
\(|\mathbf{a}|=|\mathbf{b}|\) & Isosceles triangular lattice\\
\hline
\end{tabular}
\end{center}

Without loss of generality, the \(z-\)components of vectors \(\mathbf{a}\) and \(\mathbf{b}\) are set to zero. The constant unit cell size is defined using the linear density (usually \(1/\rho_2\)). The periodicity means an integer (\(m\) and \(n\)) vector sum of \(\mathbf{a}\) and \(\mathbf{b}\) must arrive at the periodicity vector, \(\mathbf{\hat{y}}c\), defined in Fig. \ref{fig:examples}. Having two vector lengths equal forms an isosceles lattice. Solving for \(\mathbf{a}\) and \(\mathbf{b}\):
\begin{align} 
\mathbf{a}&=\frac{1}{\rho_1}\left(\begin{array}{c} -n \\ \frac{mc\rho_1-n\sqrt{c^2\rho_1^2-p^2}}{p}\end{array} \right)\\
\mathbf{b}&=\frac{1}{\rho_1}\left(\begin{array}{c} m \\ \frac{-nc\rho_1+m\sqrt{c^2\rho_1^2-p^2}}{p}\end{array} \right),
\end{align}
where \(p=m^2-n^2\). Although there are multiple solutions, they generate the lattices which are reflected on the \(x\) axis which correspond to degenerate energies and are not considered further. Note the factor of \(1/\rho_1\) and \(c\rho_1=\alpha\), where \(\alpha\) is a continuous variable. For any given value of \(\alpha\), the angles between lattice vectors are conserved, since \(\mathbf{a}\cdot\mathbf{b}/|\mathbf{a}||\mathbf{b}|\) only depends on \(\alpha\), and the absolute scale of the lattice is thus set by \(1/\rho_1\). Furthermore, the numerical transitions are consistent with \(\alpha_T=c\rho_1\) (specific values of \(\alpha\) that match up to transition curves). For example, at the \((m=2,n=2)\rightarrow (m=2,n=1)\) transition, we construct a nonlinear model of the form \(\rho_1 = \alpha/c\) which fits with \(\alpha = 6.197\) with an R-squared value of \(0.999998\). Note for this transition, the exact value of \(\alpha=8\sqrt{3/5}\simeq 6.19677\). Fitting the curves as inverse functions, we search along the line \(c=\rho_1\) and cross each transition point once (we set \(c=\rho_1=\sqrt{\alpha}\), with dimensional constants absorbed by \(\lambda\) or \(\kappa\)). At more extreme scalings, one would expect this method to break down as the details of the interactions will cause higher order effects. We then express the lattice vectors as:
\begin{align}
\mathbf{a}&=\frac{1}{\sqrt{\alpha}}\left(\begin{array}{c} -n \\ \frac{m\alpha-n\sqrt{\alpha^2-p^2}}{p}\end{array} \right)\\
\mathbf{b}&=\frac{1}{\sqrt{\alpha}}\left(\begin{array}{c} m \\ \frac{-n\alpha+m\sqrt{\alpha^2-p^2}}{p}\end{array} \right).
\end{align}
For \(\lim_{m\rightarrow n}\):
\begin{align}
\mathbf{a}=\left(\begin{array}{c} -m/\sqrt{\alpha} \\ \sqrt{\alpha}/2m\end{array} \right)
\qquad
\mathbf{b}=\left(\begin{array}{c} m/\sqrt{\alpha} \\ \ \sqrt{\alpha}/2m\end{array} \right).
\end{align}

Three primitive lattice vectors are needed for phyllotactic notation. Vectors \(\mathbf{a}\) and \(\mathbf{b}\) are primitive, the third is \(\mathbf{c^-}=\mathbf{a}-\mathbf{b}\) or \(\mathbf{c^+}=\mathbf{a}+\mathbf{b}\). \(|\mathbf{c^-}|=|\mathbf{c^+}|\) corresponds to a square lattice. This is only ever the ground state when \((m=1,n=1)\) for \(\alpha=2\), or \(\alpha_\square=m^2+n^2\). When \(\alpha<\alpha_\square\), \(\mathbf{c^+}\) is primitive and when \(\alpha>\alpha_\square\), \(\mathbf{c^-}\) is primitive. Given the primitive vectors, the accompanying phyllotactic description is \([m+n,m,n]\) when \(\alpha>\alpha_\square\) and \([m,n,m-n]\) when \(\alpha<\alpha_\square\). We use this notation alongside \((m,n)\) for clarity between our original constraints and the phyllotactic nature of the problem.

We calculate the energies generated by a set of \(m\) and \(n\) for a particular value of \(\alpha\) to find the ground state. In Appendix \ref{sec:appA}, we show the maximum value of \(m\) needed to capture the lowest energy behaviour is \(\lfloor\sqrt{2\alpha_\text{max}}/\sqrt[4]{3}\rfloor\). Where \(\alpha_\text{max}\) is the maximum value of \(\alpha\) we search with. Figure \ref{fig:sampling} shows the ground states found for \(0<\alpha<25\). By numerically searching for ground states and transitions, we are able to categorise transitions and locate degenerate points.
%%%%%%%%%%%%%%%%%%%%%%
%%  ANALYSIS %%
%%%%%%%%%%%%%%%%%%%%%%
\section{Analysis of the phenomenological model}
To explore the qualitative generality of our results we compare the numerical results for the Bessel-function and the Yukawa-potential highlighted in Fig. \ref{fig:sampling}. The largest difference occurs when the gradients of the transitioning energies are similar. This is due to the energy difference of each state being small over a larger region than the average. We will see that these transitions have a slight dependence on the interaction potential which can cause small discrepancies in transition values. Both potentials show instantaneous degenerate ground states where two states are equivalent for a critical value of \(\alpha\) but no associated structural transition is observed. 

Degenerate ground states form when, for a given value of \(\alpha(=\alpha_\triangle)\), two states both form an equilateral triangular lattice. States can be found exactly by setting \(|\mathbf{a}|=|\mathbf{a\pm b}|\). In the two-dimensional phase diagram, this corresponds to lines of \(c=\alpha_\triangle/\rho_1\) where the energy is instantaneously degenerate. We find:
\begin{align}
\alpha_\triangle=\frac{2}{\sqrt{3}}(m^2\pm mn +n^2).
\end{align}

The lower value of \(\alpha_\triangle\) only exists if \(m^2-4mn+n^2<0\). We find no transitions at \(\alpha=\alpha_\triangle\), only instantaneous degeneracies where multiple states can exist if multiple values of \(m\) and \(n\) yield the same value of \(\alpha_\triangle\).

Two types of structural transition exist:

\(\bullet\) two states sharing the exact same structure up to a global rotation or

\(\bullet\) two states with different structures that exist on opposite sides of an energetic parabola.

\begin{figure}
\centering
\includegraphics[width=8.6cm]{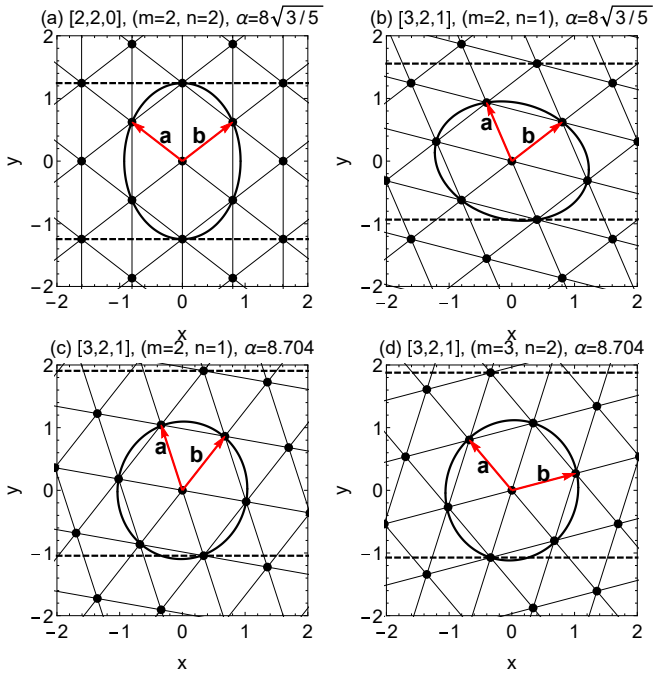}
\caption{(Color online) Renderings of states at two different transition points. Dashed lines show the repeated unit. Lattice vectors are highlighted in red. (a) and (b) \([2,2,0] (m=2, n=2)\) and \([3,2,1] (m=2, n=1)\) states at \(\alpha_T=8\sqrt{3/5}\). The states are geometrically identical aside from a global rotation. Here, \(\alpha_T\) is independent of potential. (c) and (d) \([3,2,1] (m=2, n=1)\) and \([3,2,1] (m=3, n=2)\) states at \(\alpha_T=8.704\). These states are geometrically distinct and are labelled as type \(1\) and \(5\) respectively. The circles of radius \(|\mathbf{a}|\) indicate which type of state is present, as shown in Fig. \ref{fig:generic}.}
\label{fig:statetypes}
\end{figure}
\begin{figure}
\centering
\includegraphics[width=8.6cm]{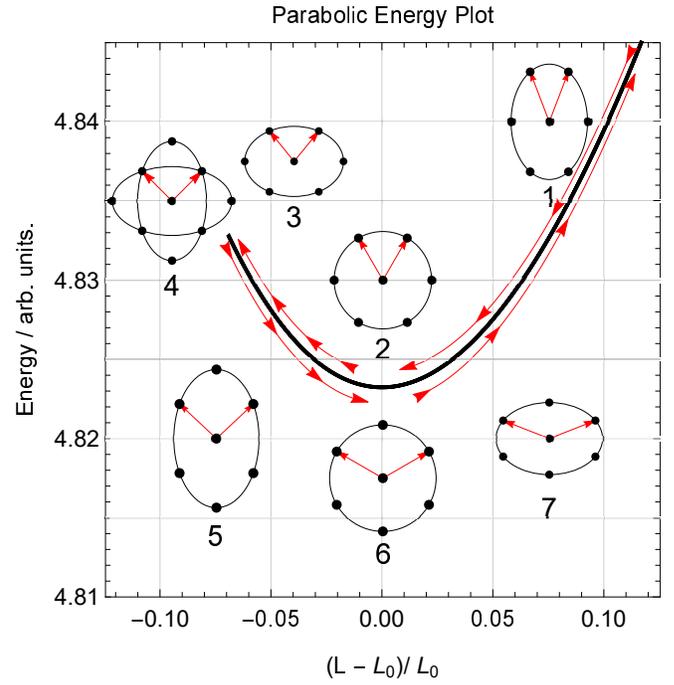}
\caption{(Color online) Plot of the energy of the generic model with respect to scaled vector length, where \(L_0=\sqrt{2/\sqrt{3}}\) is the vector length that yields the minimum energy. The numbers correspond to the region or point on the curve where certain types of lattice form. State types \(2\) and \(6\) are the perfect lattice, \(4\) is the square lattice, and \(1\), \(3\), \(5\), and \(6\) are variations that describe the relative proximity of the next-nearest neighbour to the circular radius equal to the lattice vector length. Any lattice can be categorised by checking which of the following inequalities holds for the numbered state types: \(1\): \(|\mathbf{a}|<|\mathbf{a}-\mathbf{b}|\), \(2\): \(|\mathbf{a}|=|\mathbf{a}-\mathbf{b}|\), \(3\): \(|\mathbf{a}-\mathbf{b}|<|\mathbf{a}|<|\mathbf{a}+\mathbf{b}|\), \(4\): \(|\mathbf{a}+\mathbf{b}|=|\mathbf{a}-\mathbf{b}|\), \(5\): \(|\mathbf{a}+\mathbf{b}|<|\mathbf{a}|<|\mathbf{a}-\mathbf{b}|\) \(6\): \(|\mathbf{a}|=|\mathbf{a}+\mathbf{b}|\), \(7\): \(|\mathbf{a}|>|\mathbf{a}+\mathbf{b}|\). The parabolic fit is accurate in the region near to the minimum of the curve. Supplementary video S1 shows the structure as a function of \(\gamma\) and \(L\) in enhanced clarity.}
\label{fig:generic}
\end{figure}
\begin{figure}
\includegraphics[width=8.6cm]{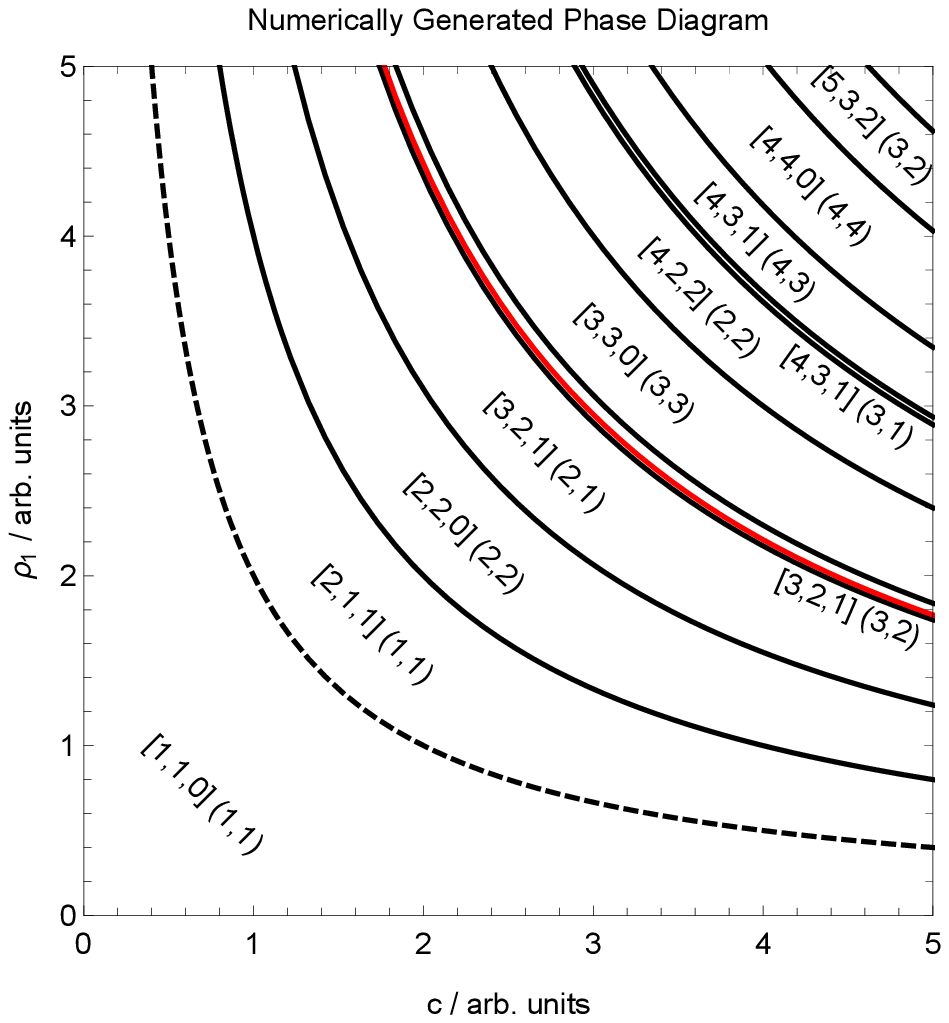}
\caption{(Color online) Rendering of the phase diagram obtained by sampling energies with the Bessel-function and Yukawa potential. Each state is labelled in accordance to the phyllotactic notation \([k+l,k,l]\) (left) and the indices from the system model, \((m,n)\) (right). The dashed line \((c=2/\rho_1)\) is where the phyllotactic notation for the \((m=1, n=1)\) state relabels itself via a unique ``relabelling'' transition. The only resolvable difference between the numerics and theoretical predictions is the position of the \([3,2,1]\) \((2, 1)\rightarrow[3,2,1]\) \((3, 2)\) transition line, with the numerical value shown in red.}
\label{fig:phasespace}
\end{figure}

%%%%%%%%%%%%%%%%%%%%%%
%% TRANSITIONS %%
%%%%%%%%%%%%%%%%%%%%%%
\section{Calculating transition points}
Figure \ref{fig:statetypes} (a) and (b) shows transitioning states when the lattices are `global rotations' of each other. The transition point where one state can be rotated into another is straightforward to describe: since all lattice vectors must be equal in length, there is an equation for \(\alpha\) given indices \(m\) and \(n\) for each state.

Figure \ref{fig:statetypes} (c) and (d) shows the more complex transition where the two lattices have different distortions. Since our analysis is along the line \(c=\rho_1\), the unit cell size is \(1\) and \(|\mathbf{a}|=|\mathbf{b}|\) still holds. Using these constraints without the periodicity needing to be satisfied, we generate a simpler description of the lattice, without any loss of generality, which is arbitrarily rotated so the chains are parallel to the horizontal axis. We parametrise this simpler lattice with \(\gamma\):
\begin{align}
\mathbf{a}=\left(\begin{array}{c} -\gamma \\1/(2\gamma) \end{array} \right) 
\qquad
\mathbf{b}=\left(\begin{array}{c} \gamma \\ 1/(2\gamma)\end{array} \right) \label{eq:gamma}.
\end{align}

This can be directly mapped to represent any of the full set of states via rotation and scaling. Figure \ref{fig:generic} illustrates the variety of lattices generated. By varying \(\gamma\) and numerically calculating ground state energies, we find a double-minima curve where the minima correspond to values of \(\alpha_\triangle\). Alternatively, if the energy is plotted against the vector length, \(L=|\mathbf{a}|=\sqrt{4\gamma^4+1}/(2\gamma)\), a parabola is found where the variation of \(\gamma\) traces out a trajectory that doubles back on itself (Supplementary Video 1 demonstrates this in detail). We can then classify the structure of each state based on its position on this parabola and lattice vector length. 

Figure \ref{fig:generic} shows the energy using Bessel function interactions, which is parabolic to leading order. Translating the curve by plotting as a function of \((L-L_0)/L_0\), where \(L_0=\sqrt{2/\sqrt{3}}\), causes it to be symmetric at the origin. \(L_0\) is the vector length corresponding to the perfect equilateral triangular lattice when \(\gamma_\triangle=1/\sqrt{2\sqrt{3}}\) or \(\gamma_\triangle=\sqrt{\sqrt{3}/2}\). The Yukawa potential similarly yields a parabolic curve but a different minimum energy value when \(L=L_0\), further indicating the qualitative consistency for different potentials.

When two equal energies are on opposite sides of the origin, we define a separation distance between points meaning the lengths of the vectors at a transition point differ by \(2d\). This leads to (\ref{eq:LTrans}) being true at the transition point. When two equal energies are on the same side of the origin, both states are the same distance away from the minimum meaning the vector lengths are equal and are expressed as (\ref{eq:LExact}). Solving either equation to calculates \(\alpha_T\) between states with indices \((k,l)\) and \((m,n)\).

\begin{align}
L(\alpha,k,l)=\sqrt{\frac{2}{\sqrt{3}}}\pm d
\qquad
L(\alpha,m,n)=\sqrt{\frac{2}{\sqrt{3}}}\mp d.
\label{eq:LTrans}
\end{align}
\begin{align}
L(\alpha,k,l)=L(\alpha,m,n).
\label{eq:LExact}
\end{align}

For (\ref{eq:LTrans}), Fig. \ref{fig:generic} indicates the signs that should be used to find \(\alpha_T\). Using the incorrect signs will yield \(d\) with the opposite sign, but the correct vector length. These equations are transcendental and can be numerically solved. It should be noted that the weak potential dependence means that these values are inexact so it is appropriate to truncate the values to a few significant figures. Figure \ref{fig:sampling} shows a slight variation in the location of some of these points as the potential is changed.

We recover the statement of equal vector lengths with (\ref{eq:LExact}), which is a special case of the parabola model. We calculate all of the exact values of \(\alpha_T\) for this case in Appendix \ref{sec:appB}.

The two-dimensional phase diagram can be numerically generated by performing the previous routine of sampling the energies of the lattice parametrised by \(c\) and \(\rho_1\). Results show that transition lines behave as expected: \(c=\alpha_T/\rho_1\). These results are then checked by finding the transition lines with the developed model. Figure \ref{fig:phasespace} shows the phase diagram for \(0<c<5\) and \(0<\rho_1<5\). Initial Monte Carlo simulations, sampling energies, and analytic methods all conform to the same picture of the phase diagram being symmetric under interchange of circumference and density.\\

TABLE II. ground states and corresponding values of \(\alpha\), capturing all transitions up to \(\alpha=25\). These results are also shown on the two-dimensional phase diagram in Fig. \ref{fig:phasespace}. Numerical values of \(\alpha_T\) are given to 2 decimal places, otherwise they are exact.
\begin{center}
\begin{tabular}{|c|c|c|}
\hline
\((m,n)\) & \([k+l,k,l]\) & \(\alpha\) \\
\hline
\((1,1)\) & \([1,1,0]\) & \(0<\alpha<2\) \\
\hline
\((1,1)\) & \([2,1,1]\) & \(2<\alpha<4\) \\
\hline
\((2,2)\) & \([2,2,0]\) & \(4<\alpha<8\sqrt{3/5}\) \\
\hline
\((2,1)\) & \([3,2,1]\) & \(8\sqrt{3/5}<\alpha<8.70\) \\
\hline
\((3,2)\) & \([3,2,1]\) & \(8.70<\alpha<9.19\) \\
\hline
\((3,3)\) & \([3,3,0]\) & \(9.19<\alpha<12\) \\
\hline
\((2,2)\) & \([4,2,2]\) & \(12<\alpha<14.45\) \\
\hline
\((3,1)\) & \([4,3,1]\) & \(14.45<\alpha<14.68\) \\
\hline
\((4,3)\) & \([4,3,1]\) & \(14.68<\alpha<16.73\) \\
\hline
\((4,4)\) & \([4,4,0]\) & \(16.73<\alpha<160/3\sqrt{7}\) \\
\hline
\((3,2)\) & \([5,3,2]\) & \(160/3\sqrt{7}<\alpha<23.10\) \\
\hline
\((5,4)\) & \([5,4,1]\) & \(23.10<\alpha<26.56\)\\
\hline
\end{tabular}
\end{center}
%%%%%%%%%%%%%%%%%%%%%%
%% REMARKS %%
%%%%%%%%%%%%%%%%%%%%%%
\section{Remarks}
Since at any value of \(\alpha_\triangle\), the corresponding state must be the ground state, we can link together these minima by searching for the transition point that must occur between them. Multiple transitions can be found between minima, so all relevant states near the minima must be considered. This can be done with confidence computationally.

We initially found ground states up to \(\alpha=25\) and \textit{correctly predicted} ground states and transitions up to \(\alpha=50\).

Similar helical structures are observed \cite{fu2017,mughal2012} in cylindrically confined systems with a general trend of increasing row numbers as one moves along the phase diagram.

We note that row transitions are a general property of confined particles. Since this system is not thermal at zero temperature, we cannot state the true order of the structural transitions. By calculating the energetic derivatives and searching for discontinuities, we find that all the transitions we observe `appear' first-order. This is typical of confined systems \cite{piacente2004}, with the only second-order transition being the zig-zag transition between one and two chains, \cite{piacente2010,straube2013}, which is absent in our system since the single chain is always unstable due to lack of global confinement.

In conclusion, we have developed a model which predicts the zero temperature ground states of identical repulsive particles confined to a cylindrical system as a function of geometry and density. Lead by a geometrical picture which emerges from the initial data, we infer an idealised description of the system and search for ground state transitions. The parabolic behaviour of the per-particle energy when measured in the reference frame of the lattice vector length allows us to write down a pair of simultaneous equations that solve for \(\alpha_T\). We find the number of rows of particles generally increases with circumference or density. The occurrences of the perfect equilateral triangular lattice divides the phase diagram into sectors that we search between in order to find transitions.

The analysis relies on the scale invariant \(\alpha\) which alone determines the ground state structure on the cylinder. This result is robust and successfully predicts the lattice structure for parameters originally out of scope of the initial phase space used to inform it. More generally, we can determine the \emph{entire} phase space for ground states in this system.

Although our analysis takes \(\alpha\) to be a global invariant, since we found these lattices forming on finite size cylinders, we predict that these structures can also form in localised regions. Taking a local value of \(\alpha\) to depend on local density and circumference, where the system could have varying circumference and density, local ground state structures might form. This result would then have a much wider application into more general systems, such as conical geometries.

The authors would like to thank J. M. F. Gunn, J. S. Watkins, J. Gartlan, R. Stanyon, H. Ansell, and C. Wilkin for their valuable insight and discussions. This work was originally presented in \cite{aatphd} and all figures are adapted from the same work. All results presented were calculated and visualised using Wolfram Mathematica. This research is funded by the EPSRC, award reference \(1366111\).
\bibliography{bibliography}

\begin{thebibliography}{21}%
\makeatletter
\providecommand \@ifxundefined [1]{%
 \@ifx{#1\undefined}
}%
\providecommand \@ifnum [1]{%
 \ifnum #1\expandafter \@firstoftwo
 \else \expandafter \@secondoftwo
 \fi
}%
\providecommand \@ifx [1]{%
 \ifx #1\expandafter \@firstoftwo
 \else \expandafter \@secondoftwo
 \fi
}%
\providecommand \natexlab [1]{#1}%
\providecommand \enquote  [1]{``#1''}%
\providecommand \bibnamefont  [1]{#1}%
\providecommand \bibfnamefont [1]{#1}%
\providecommand \citenamefont [1]{#1}%
\providecommand \href@noop [0]{\@secondoftwo}%
\providecommand \href [0]{\begingroup \@sanitize@url \@href}%
\providecommand \@href[1]{\@@startlink{#1}\@@href}%
\providecommand \@@href[1]{\endgroup#1\@@endlink}%
\providecommand \@sanitize@url [0]{\catcode `\\12\catcode `\$12\catcode
  `\&12\catcode `\#12\catcode `\^12\catcode `\_12\catcode `\%12\relax}%
\providecommand \@@startlink[1]{}%
\providecommand \@@endlink[0]{}%
\providecommand \url  [0]{\begingroup\@sanitize@url \@url }%
\providecommand \@url [1]{\endgroup\@href {#1}{\urlprefix }}%
\providecommand \urlprefix  [0]{URL }%
\providecommand \Eprint [0]{\href }%
\providecommand \doibase [0]{http://dx.doi.org/}%
\providecommand \selectlanguage [0]{\@gobble}%
\providecommand \bibinfo  [0]{\@secondoftwo}%
\providecommand \bibfield  [0]{\@secondoftwo}%
\providecommand \translation [1]{[#1]}%
\providecommand \BibitemOpen [0]{}%
\providecommand \bibitemStop [0]{}%
\providecommand \bibitemNoStop [0]{.\EOS\space}%
\providecommand \EOS [0]{\spacefactor3000\relax}%
\providecommand \BibitemShut  [1]{\csname bibitem#1\endcsname}%
\let\auto@bib@innerbib\@empty
%</preamble>
\bibitem [{\citenamefont {Lohr}\ \emph {et~al.}(2010)\citenamefont {Lohr},
  \citenamefont {Alsayed}, \citenamefont {Chen}, \citenamefont {Zhang},
  \citenamefont {Kamien},\ and\ \citenamefont {Yodh}}]{lohr2010}%
  \BibitemOpen
  \bibfield  {author} {\bibinfo {author} {\bibfnamefont {M.~A.}\ \bibnamefont
  {Lohr}}, \bibinfo {author} {\bibfnamefont {A.~M.}\ \bibnamefont {Alsayed}},
  \bibinfo {author} {\bibfnamefont {B.~G.}\ \bibnamefont {Chen}}, \bibinfo
  {author} {\bibfnamefont {Z.}~\bibnamefont {Zhang}}, \bibinfo {author}
  {\bibfnamefont {R.~D.}\ \bibnamefont {Kamien}}, \ and\ \bibinfo {author}
  {\bibfnamefont {A.~G.}\ \bibnamefont {Yodh}},\ }\href {\doibase
  10.1103/physreve.81.040401} {\bibfield  {journal} {\bibinfo  {journal} {Phys.
  Rev. E}\ }\textbf {\bibinfo {volume} {81}},\ \bibinfo {pages} {040401(R)}
  (\bibinfo {year} {2010})}\BibitemShut {NoStop}%
\bibitem [{\citenamefont {Mughal}\ \emph {et~al.}(2011)\citenamefont {Mughal},
  \citenamefont {Chan},\ and\ \citenamefont {Weaire}}]{mughal2011}%
  \BibitemOpen
  \bibfield  {author} {\bibinfo {author} {\bibfnamefont {A.}~\bibnamefont
  {Mughal}}, \bibinfo {author} {\bibfnamefont {H.~K.}\ \bibnamefont {Chan}}, \
  and\ \bibinfo {author} {\bibfnamefont {D.}~\bibnamefont {Weaire}},\ }\href
  {\doibase 10.1103/physrevlett.106.115704} {\bibfield  {journal} {\bibinfo
  {journal} {Phys. Rev. Lett.}\ }\textbf {\bibinfo {volume} {106}},\ \bibinfo
  {pages} {115704} (\bibinfo {year} {2011})}\BibitemShut {NoStop}%
\bibitem [{\citenamefont {Mughal}\ and\ \citenamefont
  {Weaire}(2017)}]{mughal2017}%
  \BibitemOpen
  \bibfield  {author} {\bibinfo {author} {\bibfnamefont {A.}~\bibnamefont
  {Mughal}}\ and\ \bibinfo {author} {\bibfnamefont {D.}~\bibnamefont
  {Weaire}},\ }\href {\doibase 10.1103/PhysRevE.95.022401} {\bibfield
  {journal} {\bibinfo  {journal} {Phys. Rev. E}\ }\textbf {\bibinfo {volume}
  {95}},\ \bibinfo {pages} {022401} (\bibinfo {year} {2017})}\BibitemShut
  {NoStop}%
\bibitem [{\citenamefont {Erickson}(1973)}]{erickson1973}%
  \BibitemOpen
  \bibfield  {author} {\bibinfo {author} {\bibfnamefont {R.~O.}\ \bibnamefont
  {Erickson}},\ }\href {\doibase 10.1126/science.181.4101.705} {\bibfield
  {journal} {\bibinfo  {journal} {Science}\ }\textbf {\bibinfo {volume}
  {181}},\ \bibinfo {pages} {705} (\bibinfo {year} {1973})}\BibitemShut
  {NoStop}%
\bibitem [{\citenamefont {Godfrey}\ and\ \citenamefont
  {Moore}(2014)}]{godfrey2014}%
  \BibitemOpen
  \bibfield  {author} {\bibinfo {author} {\bibfnamefont {M.~J.}\ \bibnamefont
  {Godfrey}}\ and\ \bibinfo {author} {\bibfnamefont {M.~A.}\ \bibnamefont
  {Moore}},\ }\href {\doibase 10.1103/physreve.89.032111} {\bibfield  {journal}
  {\bibinfo  {journal} {Phys. Rev. E}\ }\textbf {\bibinfo {volume} {89}},\
  \bibinfo {pages} {032111} (\bibinfo {year} {2014})}\BibitemShut {NoStop}%
\bibitem [{\citenamefont {Kofke}\ and\ \citenamefont {Post}(1993)}]{kofke1993}%
  \BibitemOpen
  \bibfield  {author} {\bibinfo {author} {\bibfnamefont {D.~A.}\ \bibnamefont
  {Kofke}}\ and\ \bibinfo {author} {\bibfnamefont {A.~J.}\ \bibnamefont
  {Post}},\ }\href {\doibase 10.1063/1.464967} {\bibfield  {journal} {\bibinfo
  {journal} {J. Chem. Phys.}\ }\textbf {\bibinfo {volume} {98}},\ \bibinfo
  {pages} {4853} (\bibinfo {year} {1993})}\BibitemShut {NoStop}%
\bibitem [{\citenamefont {Khlobystov}\ \emph {et~al.}(2004)\citenamefont
  {Khlobystov}, \citenamefont {Britz}, \citenamefont {Ardavan},\ and\
  \citenamefont {Briggs}}]{khlobystov2004}%
  \BibitemOpen
  \bibfield  {author} {\bibinfo {author} {\bibfnamefont {A.~N.}\ \bibnamefont
  {Khlobystov}}, \bibinfo {author} {\bibfnamefont {D.~A.}\ \bibnamefont
  {Britz}}, \bibinfo {author} {\bibfnamefont {A.}~\bibnamefont {Ardavan}}, \
  and\ \bibinfo {author} {\bibfnamefont {G.~A.~D.}\ \bibnamefont {Briggs}},\
  }\href {\doibase 10.1103/physrevlett.92.245507} {\bibfield  {journal}
  {\bibinfo  {journal} {Phys. Rev. Lett.}\ }\textbf {\bibinfo {volume} {92}},\
  \bibinfo {pages} {245507} (\bibinfo {year} {2004})}\BibitemShut {NoStop}%
\bibitem [{\citenamefont {Liang}\ \emph {et~al.}(2014)\citenamefont {Liang},
  \citenamefont {Xu}, \citenamefont {Deng}, \citenamefont {Wang}, \citenamefont
  {Liu}, \citenamefont {Li},\ and\ \citenamefont {Zhu}}]{liang2014}%
  \BibitemOpen
  \bibfield  {author} {\bibinfo {author} {\bibfnamefont {R.}~\bibnamefont
  {Liang}}, \bibinfo {author} {\bibfnamefont {J.}~\bibnamefont {Xu}}, \bibinfo
  {author} {\bibfnamefont {R.}~\bibnamefont {Deng}}, \bibinfo {author}
  {\bibfnamefont {K.}~\bibnamefont {Wang}}, \bibinfo {author} {\bibfnamefont
  {S.}~\bibnamefont {Liu}}, \bibinfo {author} {\bibfnamefont {J.}~\bibnamefont
  {Li}}, \ and\ \bibinfo {author} {\bibfnamefont {J.}~\bibnamefont {Zhu}},\
  }\href {\doibase 10.1021/mz5002146} {\bibfield  {journal} {\bibinfo
  {journal} {ACS Macro Lett}\ }\textbf {\bibinfo {volume} {3}},\ \bibinfo
  {pages} {486} (\bibinfo {year} {2014})}\BibitemShut {NoStop}%
\bibitem [{\citenamefont {Yamazaki}\ \emph {et~al.}(2008)\citenamefont
  {Yamazaki}, \citenamefont {Kuramochi}, \citenamefont {Takagi}, \citenamefont
  {Homma}, \citenamefont {Nishimura}, \citenamefont {Hori}, \citenamefont
  {Watanabe}, \citenamefont {Suzuki},\ and\ \citenamefont
  {Kobayashi}}]{yamazaki2008}%
  \BibitemOpen
  \bibfield  {author} {\bibinfo {author} {\bibfnamefont {T.}~\bibnamefont
  {Yamazaki}}, \bibinfo {author} {\bibfnamefont {K.}~\bibnamefont {Kuramochi}},
  \bibinfo {author} {\bibfnamefont {D.}~\bibnamefont {Takagi}}, \bibinfo
  {author} {\bibfnamefont {Y.}~\bibnamefont {Homma}}, \bibinfo {author}
  {\bibfnamefont {F.}~\bibnamefont {Nishimura}}, \bibinfo {author}
  {\bibfnamefont {N.}~\bibnamefont {Hori}}, \bibinfo {author} {\bibfnamefont
  {K.}~\bibnamefont {Watanabe}}, \bibinfo {author} {\bibfnamefont
  {S.}~\bibnamefont {Suzuki}}, \ and\ \bibinfo {author} {\bibfnamefont
  {Y.}~\bibnamefont {Kobayashi}},\ }\href {\doibase
  10.1088/0957-4484/19/04/045702} {\bibfield  {journal} {\bibinfo  {journal}
  {Nanotechnology}\ }\textbf {\bibinfo {volume} {19}},\ \bibinfo {pages}
  {045702} (\bibinfo {year} {2008})}\BibitemShut {NoStop}%
\bibitem [{\citenamefont {Mughal}\ and\ \citenamefont
  {Weaire}(2014)}]{mughal2013}%
  \BibitemOpen
  \bibfield  {author} {\bibinfo {author} {\bibfnamefont {A.}~\bibnamefont
  {Mughal}}\ and\ \bibinfo {author} {\bibfnamefont {D.}~\bibnamefont
  {Weaire}},\ }\href {\doibase 10.1103/PhysRevE.89.042307} {\bibfield
  {journal} {\bibinfo  {journal} {Phys. Rev. E}\ }\textbf {\bibinfo {volume}
  {89}},\ \bibinfo {pages} {042307} (\bibinfo {year} {2014})}\BibitemShut
  {NoStop}%
\bibitem [{\citenamefont {Airy}\ and\ \citenamefont {Darwin}(1873)}]{airy1872}%
  \BibitemOpen
  \bibfield  {author} {\bibinfo {author} {\bibfnamefont {H.}~\bibnamefont
  {Airy}}\ and\ \bibinfo {author} {\bibfnamefont {C.~R.}\ \bibnamefont
  {Darwin}},\ }\href {\doibase 10.1098/rspl.1872.0040} {\bibfield  {journal}
  {\bibinfo  {journal} {Proc. R. Soc. Lond.}\ }\textbf {\bibinfo {volume}
  {21}},\ \bibinfo {pages} {176} (\bibinfo {year} {1873})}\BibitemShut
  {NoStop}%
\bibitem [{\citenamefont {Fu}\ \emph {et~al.}(2016)\citenamefont {Fu},
  \citenamefont {Steinhardt}, \citenamefont {Zhao}, \citenamefont {Socolar},\
  and\ \citenamefont {Charbonneau}}]{fu2016}%
  \BibitemOpen
  \bibfield  {author} {\bibinfo {author} {\bibfnamefont {L.}~\bibnamefont
  {Fu}}, \bibinfo {author} {\bibfnamefont {W.}~\bibnamefont {Steinhardt}},
  \bibinfo {author} {\bibfnamefont {H.}~\bibnamefont {Zhao}}, \bibinfo {author}
  {\bibfnamefont {J.~E.~S.}\ \bibnamefont {Socolar}}, \ and\ \bibinfo {author}
  {\bibfnamefont {P.}~\bibnamefont {Charbonneau}},\ }\href {\doibase
  10.1039/c5sm02875b} {\bibfield  {journal} {\bibinfo  {journal} {Soft Matter}\
  }\textbf {\bibinfo {volume} {12}},\ \bibinfo {pages} {2505} (\bibinfo {year}
  {2016})}\BibitemShut {NoStop}%
\bibitem [{\citenamefont {Beller}\ and\ \citenamefont
  {Nelson}(2016)}]{nelson2016}%
  \BibitemOpen
  \bibfield  {author} {\bibinfo {author} {\bibfnamefont {D.~A.}\ \bibnamefont
  {Beller}}\ and\ \bibinfo {author} {\bibfnamefont {D.~R.}\ \bibnamefont
  {Nelson}},\ }\href {\doibase 10.1103/PhysRevE.94.033004} {\bibfield
  {journal} {\bibinfo  {journal} {Phys. Rev. E}\ }\textbf {\bibinfo {volume}
  {94}},\ \bibinfo {pages} {033004} (\bibinfo {year} {2016})}\BibitemShut
  {NoStop}%
\bibitem [{\citenamefont {Piacente}\ \emph {et~al.}(2004)\citenamefont
  {Piacente}, \citenamefont {Schweigert}, \citenamefont {Betouras},\ and\
  \citenamefont {Peeters}}]{piacente2004}%
  \BibitemOpen
  \bibfield  {author} {\bibinfo {author} {\bibfnamefont {G.}~\bibnamefont
  {Piacente}}, \bibinfo {author} {\bibfnamefont {I.~V.}\ \bibnamefont
  {Schweigert}}, \bibinfo {author} {\bibfnamefont {J.~J.}\ \bibnamefont
  {Betouras}}, \ and\ \bibinfo {author} {\bibfnamefont {F.~M.}\ \bibnamefont
  {Peeters}},\ }\href {\doibase 10.1103/physrevb.69.045324} {\bibfield
  {journal} {\bibinfo  {journal} {Phys. Rev. B}\ }\textbf {\bibinfo {volume}
  {69}},\ \bibinfo {pages} {045324} (\bibinfo {year} {2004})}\BibitemShut
  {NoStop}%
\bibitem [{\citenamefont {O\u{g}uz}\ \emph {et~al.}(2011)\citenamefont
  {O\u{g}uz}, \citenamefont {Messina},\ and\ \citenamefont {L{\"
  o}wen}}]{oguz2011}%
  \BibitemOpen
  \bibfield  {author} {\bibinfo {author} {\bibfnamefont {E.~C.}\ \bibnamefont
  {O\u{g}uz}}, \bibinfo {author} {\bibfnamefont {R.}~\bibnamefont {Messina}}, \
  and\ \bibinfo {author} {\bibfnamefont {H.}~\bibnamefont {L{\" o}wen}},\
  }\href {\doibase 10.1209/0295-5075/94/28005} {\bibfield  {journal} {\bibinfo
  {journal} {EPL}\ }\textbf {\bibinfo {volume} {94}},\ \bibinfo {pages} {28005}
  (\bibinfo {year} {2011})}\BibitemShut {NoStop}%
\bibitem [{\citenamefont {Amir}\ \emph {et~al.}(2013)\citenamefont {Amir},
  \citenamefont {Paulose},\ and\ \citenamefont {Nelson}}]{amir2013}%
  \BibitemOpen
  \bibfield  {author} {\bibinfo {author} {\bibfnamefont {A.}~\bibnamefont
  {Amir}}, \bibinfo {author} {\bibfnamefont {J.}~\bibnamefont {Paulose}}, \
  and\ \bibinfo {author} {\bibfnamefont {D.~R.}\ \bibnamefont {Nelson}},\
  }\href {\doibase 10.1103/physreve.87.042314} {\bibfield  {journal} {\bibinfo
  {journal} {Phys. Rev. E}\ }\textbf {\bibinfo {volume} {87}},\ \bibinfo
  {pages} {042314} (\bibinfo {year} {2013})}\BibitemShut {NoStop}%
\bibitem [{\citenamefont {Fu}\ \emph {et~al.}(2017)\citenamefont {Fu},
  \citenamefont {Bian}, \citenamefont {Shields}, \citenamefont {Cruz},
  \citenamefont {L\'{o}pez},\ and\ \citenamefont {Charbonneau}}]{fu2017}%
  \BibitemOpen
  \bibfield  {author} {\bibinfo {author} {\bibfnamefont {L.}~\bibnamefont
  {Fu}}, \bibinfo {author} {\bibfnamefont {C.}~\bibnamefont {Bian}}, \bibinfo
  {author} {\bibfnamefont {C.~W.}\ \bibnamefont {Shields}}, \bibinfo {author}
  {\bibfnamefont {D.~F.}\ \bibnamefont {Cruz}}, \bibinfo {author}
  {\bibfnamefont {G.~P.}\ \bibnamefont {L\'{o}pez}}, \ and\ \bibinfo {author}
  {\bibfnamefont {P.}~\bibnamefont {Charbonneau}},\ }\href {\doibase
  10.1039/C7SM00316A} {\bibfield  {journal} {\bibinfo  {journal} {Soft Matter}\
  }\textbf {\bibinfo {volume} {13}},\ \bibinfo {pages} {3296} (\bibinfo {year}
  {2017})}\BibitemShut {NoStop}%
\bibitem [{\citenamefont {Mughal}\ \emph {et~al.}(2012)\citenamefont {Mughal},
  \citenamefont {Chan}, \citenamefont {Weaire},\ and\ \citenamefont
  {Hutzler}}]{mughal2012}%
  \BibitemOpen
  \bibfield  {author} {\bibinfo {author} {\bibfnamefont {A.}~\bibnamefont
  {Mughal}}, \bibinfo {author} {\bibfnamefont {H.~K.}\ \bibnamefont {Chan}},
  \bibinfo {author} {\bibfnamefont {D.}~\bibnamefont {Weaire}}, \ and\ \bibinfo
  {author} {\bibfnamefont {S.}~\bibnamefont {Hutzler}},\ }\href {\doibase
  10.1103/physreve.85.051305} {\bibfield  {journal} {\bibinfo  {journal} {Phys.
  Rev. E}\ }\textbf {\bibinfo {volume} {85}},\ \bibinfo {pages} {051305}
  (\bibinfo {year} {2012})}\BibitemShut {NoStop}%
\bibitem [{\citenamefont {Piacente}\ \emph {et~al.}(2010)\citenamefont
  {Piacente}, \citenamefont {Hai},\ and\ \citenamefont
  {Peeters}}]{piacente2010}%
  \BibitemOpen
  \bibfield  {author} {\bibinfo {author} {\bibfnamefont {G.}~\bibnamefont
  {Piacente}}, \bibinfo {author} {\bibfnamefont {G.~Q.}\ \bibnamefont {Hai}}, \
  and\ \bibinfo {author} {\bibfnamefont {F.~M.}\ \bibnamefont {Peeters}},\
  }\href {\doibase 10.1103/physrevb.81.024108} {\bibfield  {journal} {\bibinfo
  {journal} {Phys. Rev. B}\ }\textbf {\bibinfo {volume} {81}},\ \bibinfo
  {pages} {024108} (\bibinfo {year} {2010})}\BibitemShut {NoStop}%
\bibitem [{\citenamefont {Straube}\ \emph {et~al.}(2013)\citenamefont
  {Straube}, \citenamefont {Dullens}, \citenamefont {Schimansky-Geier},\ and\
  \citenamefont {Louis}}]{straube2013}%
  \BibitemOpen
  \bibfield  {author} {\bibinfo {author} {\bibfnamefont {A.~V.}\ \bibnamefont
  {Straube}}, \bibinfo {author} {\bibfnamefont {R.~P.~A.}\ \bibnamefont
  {Dullens}}, \bibinfo {author} {\bibfnamefont {L.}~\bibnamefont
  {Schimansky-Geier}}, \ and\ \bibinfo {author} {\bibfnamefont {A.~A.}\
  \bibnamefont {Louis}},\ }\href {\doibase 10.1063/1.4823501} {\bibfield
  {journal} {\bibinfo  {journal} {J. Chem. Phys.}\ }\textbf {\bibinfo {volume}
  {139}},\ \bibinfo {pages} {134908} (\bibinfo {year} {2013})}\BibitemShut
  {NoStop}%
\bibitem [{\citenamefont {Tomlinson}(2018)}]{aatphd}%
  \BibitemOpen
  \bibfield  {author} {\bibinfo {author} {\bibfnamefont {A.~A.}\ \bibnamefont
  {Tomlinson}},\ }\href@noop {} {Ph.D. thesis},\ \bibinfo  {school} {University
  of Birmingham} (\bibinfo {year} {2018})\BibitemShut {NoStop}%
\end{thebibliography}%

%%%%%%%%%%%%%%%%%%%%%%
%% APPENDIX %%
%%%%%%%%%%%%%%%%%%%%%%
\appendix
\section{Determining \(m_{\text{max}}\)}\label{sec:appA}
When calculating state energies numerically, it is useful to have a sensible number of states defined by \(m\) and \(n\) to search through. Since \(m \ge n\), we need to determine the largest value of \(m\) only. This is determined by considering when a combination of \(m\) and \(n\) is likely to describe a ground state in the vicinity of \(\alpha_\triangle\). For a particular choice of \(m\), a ground state can be first yielded near
\begin{align}\label{eq:alphatrimin}
\alpha = 2(m^2 - mn + n^2)/\sqrt{3}.
\end{align}

Minimising with respect to \(n\) gives:
\begin{align}
\frac{\partial \alpha}{\partial n} = \frac{2}{\sqrt{3}}\left(2n - m\right) = 0,
\end{align}
which is optimised with \(n = m/2\) (ignoring half integers). Substituting back into (\ref{eq:alphatrimin}) gives the result:
\begin{align}
m = \sqrt{\frac{2 \alpha}{\sqrt{3}}}.
\end{align}

This is then rounded down to the nearest integer.

\section{Exact values of \(\alpha_T\)}\label{sec:appB}
When two states at a transition point share the same structure up to a global rotation, we can solve exactly for the transition points. These solutions are independent of interaction potential and solely dependent on geometry.

For this type of transition between states with indices \((k,l)\) and \((m,n)\), we calculate \(\alpha_T\) and determine physical validity. Since lattice vectors lengths are constrained to be equal, we formulate \(|\mathbf{a}(m,n)|=|\mathbf{a}(k,l)|\) which leads to:
\begin{align}\label{eq:master}
\begin{split}
0=\alpha\left(\frac{\left(k^2+l^2\right)}{q^2}-\frac{\left(m^2+n^2\right)}{p^2}\right)\\
-\frac{2kl\sqrt{\alpha^2-q^2}}{q^2}+\frac{2mn\sqrt{\alpha^2-p^2}}{p^2},
\end{split}
\end{align}
where \(q=k^2-l^2\). This can be manipulated into a quadratic equation in \(\alpha^2\) that yields four solutions shown in (\ref{eq:alpha1}) and (\ref{eq:alpha2}).
\begin{widetext}
\begin{align}
\label{eq:alpha1}
\alpha_1^\pm=\pm\frac{2(lm+kn)(km+ln)}{\sqrt{-(k-l-m-n)(k+l+m-n)(k+l-m+n)(k-l+m+n)}}
\end{align}
\begin{align}
\label{eq:alpha2}
\alpha_2^\pm=\pm\frac{2(lm-kn)(km-ln)}{\sqrt{-(k+l-m-n)(k-l+m-n)(k-l-m+n)(k+l+m+n)}}
\end{align}
\end{widetext}
The solutions are only valid if they also satisfy (\ref{eq:master}). Indices alone can determine validity:

\(\bullet\) \(\alpha_1^+\) is valid if \(-4klmn+f_1<0\) and \(4klmn-f_2>0\).

\(\bullet\) \(\alpha_1^-\) is valid if \(-4klmn+f_1<0\) and \(4klmn-f_2<0\).

\(\bullet\) \(\alpha_2^+\) is valid if \(-4klmn-f_2<0\) and \(4klmn+f_1<0\).

\(\bullet\) \(\alpha_2^-\) is valid if \(-4klmn-f_2>0\) and \(4klmn+f_1>0\).

Here, \(f_1(k,l,m,n)=q^2-(k^2+l^2)(m^2+n^2)\) and \(f_2(k,l,m,n)=p^2-(k^2+l^2)(m^2+n^2)\). If a valid value of \(\alpha\) is real and positive, then it is a true physical solution for the transition between \(N_{k,l}\) and \(N^{\prime}_{m,n}\).
\end{document}